\documentclass[conference]{IEEEtran}
\IEEEoverridecommandlockouts
\usepackage{cite}
\usepackage{amsmath,amssymb,amsfonts}
\usepackage{algorithmic}
\usepackage{graphicx}
\usepackage{subcaption}  
\usepackage{textcomp}
\usepackage{xcolor, enumitem}
\def\BibTeX{{\rm B\kern-.05em{\sc i\kern-.025em b}\kern-.08em
    T\kern-.1667em\lower.7ex\hbox{E}\kern-.125emX}}
\begin{document}
\newcommand{\am}[1]{\textcolor{red}{#1}}
\setlist[itemize]{leftmargin=*} 

\title{Class-Incremental Learning for Sound Event Localization and Detection\\
\thanks{This work was supported by Jane and Aatos Erkko Foundation under grant number 230048, "Continual learning of sounds with deep neural networks". \\
The authors wish to thank CSC-IT Centre of Science Ltd., Finland,  for providing computational resources.}
}

\author{Ruchi Pandey, Manjunath Mulimani, Archontis Politis, and Annamaria Mesaros\\
Signal Processing Research Centre, Tampere University, Finland
}


\maketitle
\begin{abstract}
 This paper investigates the feasibility of class-incremental learning (CIL) for Sound Event Localization and Detection (SELD) tasks. The method features an incremental learner that can learn new sound classes independently while preserving knowledge of old classes. The continual learning is achieved through a mean square error-based distillation loss to minimize output discrepancies between subsequent learners. The experiments are conducted on the TAU-NIGENS Spatial Sound Events 2021 dataset, which includes 12 different sound classes and demonstrate the efficacy of proposed method. We begin by learning 8 classes and introduce the 4 new classes at next stage. After the incremental phase, the system is evaluated on the full set of learned classes. Results show that, for this realistic dataset, our proposed method successfully maintains baseline performance across all metrics.
 
\end{abstract}


\section{Introduction}
Sound Event Localization and Detection (SELD) is an array signal processing task focused on simultaneously detecting the temporal occurrence of target sound events and localizing their spatial positions when they are active \cite{adavanne2018direction}. It is a critical task in various practical applications such as surveillance, robotics, and smart home devices, where accurately identifying and locating sound events is important for decision-making and situational awareness \cite{adavanne2018sound,wan2016application,farmani2015maximum,grumiaux2022survey}. 
 While deep learning approaches have significantly advanced SELD's detection and localization precision \cite{adavanne2018direction, shimada2021accdoa, mesaros2019joint, cao2021improved}, these models are typically trained on a fixed set of sound classes. For practical, low-cost array processing, real-world systems need the flexibility to add new classes without retraining on all previous data. This limitation in adaptability motivates the need for methods that can effectively incorporate new classes into existing models without compromising performance on previously learned classes.


A common strategy to improve the performance of pretrained models across new sound classes is fine-tuning, where a model initially trained on a set of classes is further trained on datasets containing new classes \cite{jung2019polyphonic}. While this approach can facilitate the incorporation of new information, it often leads to catastrophic forgetting, a phenomenon that occurs when the performance of the model on previously learned classes deteriorates as the model is trained on new classes. This challenge underscores the need for models that can expand their capabilities over time without sacrificing the accuracy on earlier learned classes \cite{xu2023semi,xiao2024dual}.

Continual learning offers a promising solution to this problem by enabling models to incrementally learn new tasks while preserving previously acquired knowledge \cite{hou2019learning,parisi2019continual}. In particular, Class-Incremental Learning (CIL) allows models to sequentially incorporate new classes into their learning architecture \cite{rebuffi2017icarl,van2019three}. CIL utilizes an expanding classifier, allowing the model to grow as new classes become available without the need for full retraining \cite{li2017learning}. This method has proven effective in fields like computer vision \cite{garg2022multi}, natural language processing \cite{bhatt2024characterizing}, and audio related tasks like acoustic scene classification and keyword spotting \cite{mulimani2024class,mulimani2023incremental,huang2022progressive,xiao2022continual}. However, only a single study to date has explored continual learning for sound source localization \cite{xiao2024configurable}, and no existing CIL work addresses the spatial analysis of complex acoustic scenes using microphone arrays. Our approach tackles the more comprehensive task of both sound event detection and localization, advancing the application of CIL for SELD.

In this paper, we propose a method for continual learning for SELD, which extends the capabilities of SELD models by incorporating CIL. The model begins with a base set of 8 sound classes, followed by an incremental stage where new sound classes are introduced. This approach is efficient as it reduces computational costs compared to retraining the model from scratch each time the set of classes changes or grows. The CIL-SELD model employs a mean square error (MSE) based distillation loss, which minimizes discrepancies between the outputs corresponding to the previously learned classes in successive learners. This enables the model to retain knowledge of earlier classes while learning the new one, facilitating recognition of classes learned at different stages in the training. 

The main contribution of this work are as follows:
\begin{itemize}
    \item We investigate CIL-SELD for incorporating class incremental learning into SELD models, enabling adaptive learning of new sound classes while retaining previously acquired knowledge. To the best of our knowledge, this is the first study on integrating incremental learning into SELD.
    \item We perform an ablation on the type of distillation losses and combination of losses in training, to understand the effect of different parameters in the incremental learning process.
\end{itemize}


\section{Class-Incremental Learning for SELD}
\label{sec:cl}

\begin{figure} 
   \centering
	\includegraphics[width=0.49\textwidth]{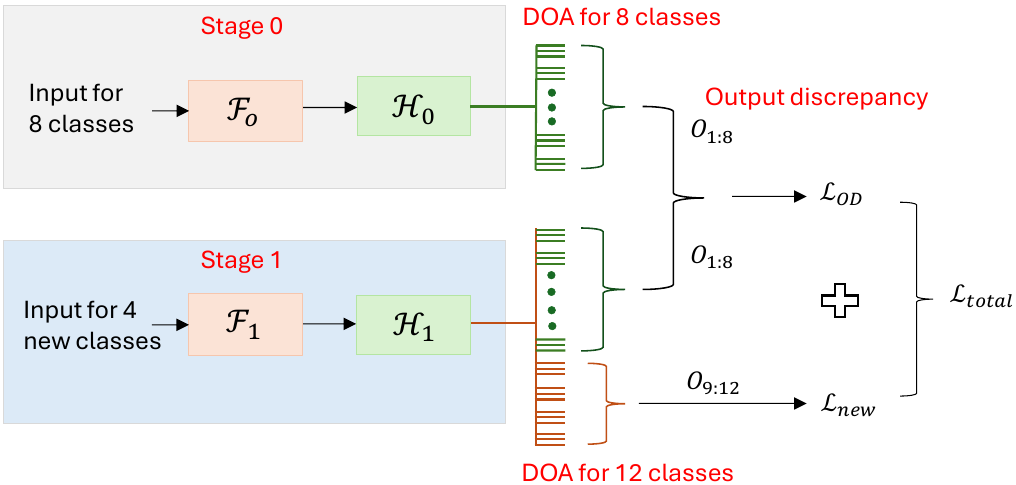}
	\caption{ Two stage class-incremental learning for SELD.}
	\label{fig:blk_diag}
\vspace{-0.4cm}
\end{figure}

\textbf{SELD:} 
To perform sound event detection and localization, a signal comprising a mixture of multiple sound events originating from various spatial locations is modeled as a multi-output regression task. In this setup, the Activity-Coupled Cartesian DOA (ACCDOA) format is used to jointly represent sound event detection (SED) and direction of arrival (DOA) \cite{shimada2021accdoa}. ACCDOA provides both the activation status of sound events and their spatial localization by encoding each active sound event with three coordinates (x, y, z) representing its DOA. When the vector magnitude formed by these coordinates exceeds a threshold, the event is considered active, and the corresponding DOA coordinates are used as its predicted DOA.
We use a convolutional recurrent neural network (CRNN) architecture based on the DCASE 2021 baseline \cite{politis2021dataset}. The model is trained on log-mel spectrograms and acoustic intensity vectors extracted from multichannel audio. By analyzing consecutive feature frames, the CRNN predicts active sound events and their spatial positions. The model outputs a single ACCDOA sequence, encoding both SED and DOA information, which allows it to localize and detect sound events in complex acoustic environments. Further details on the SELD architecture can be found in \cite{politis2021dataset}.

\textbf{Class-incremental learning for SELD:} Figure 1 illustrates a two-stage incremental framework for SELD. In \textbf{Stage 0}, the SELD model is trained for 8 sound classes. The network consists of two main components: a feature extractor, denoted as $\mathcal{F}_0$, which produces high-level data representations, and a regression model, $\mathcal{H}_0$, with 8x3 output neurons corresponding to the ACCDOA representation of the 8 sound classes. In \textbf{Stage 1}, the regression model $\mathcal{H}_0$ is expanded to 12 classes by adding 4 new classes to obtain the updated regression model $\mathcal{H}_1$. This expansion adds 12 neurons to represent the ACCDOA outputs for the 4 new classes. The feature extractor $\mathcal{F}_1$ remains the same as $\mathcal{F}_0$ but will continue to be trained during Stage 1 using the new data. 

A significant challenge in this phase is catastrophic forgetting, where the model risks losing its ability to accurately recognize the original 8 classes while learning the new ones. To address catastrophic forgetting, we employ an output distillation loss ($\mathcal{L}_{\mathcal{OD}}$), similar to \cite{mulimani2024class}. We use a mean square error (MSE) distillation loss to align it with the general training procedure for SELD. This loss measures the discrepancy between the outputs of the original 8 classes, as predicted by the model at the previous stage, and the outputs predicted by the updated model when presented with the same input. By minimizing this discrepancy, the model retains a similar behavior on the original 8 classes while learning to detect and localize the 4 new classes. The total loss function for CIL-SELD, which balances between retaining old knowledge and acquiring new information, is defined as follows:
 \begin{equation}\label{eq:1}
     \mathcal{L} = (1 - \lambda) \mathcal{L}_{\text{MSE}} + \lambda ~ \mathcal{L}_{\mathcal{OD}}
 \end{equation}
where $\mathcal{L}_{\text{MSE}}$ represents the MSE loss between the target and predicted outputs for the 4 new classes introduced in Stage 1, and $\mathcal{L}_{\mathcal{OD}}$ is the distillation loss, which measures the difference between the outputs from the frozen old model (Stage 0) and the outputs predicted by the updated model in Stage 1 for the original 8 classes. The parameter $\lambda$ regulates the trade-off between learning the new classes and preserving knowledge on the old classes.
\section{Experimental Setup}
\label{sec:results}
\subsection{Dataset}
We used the TAU-NIGENS Spatial Sound Events 2021 dataset containing spatial sound-scene recordings that consist of sound events of different categories integrated into a variety of acoustical spaces, from multiple directions and distances \cite{politis2021dataset}. Apart from the spatialized sound events of the target classes, sound events not belonging to any of the target classes are also included in the scene. The development dataset contains 600 one-minute long sound scene recordings sampled at 24kHz. We used first-order ambisonics (FOA) format from a tetrahedral microphone array. The 12 target sound classes of the spatialized events are: \emph{female scream, female speech, footsteps, knocking on door, male scream, male speech, ringing phone, piano, alarm, crying baby, crash, barking dog}. These classes are assigned IDs 1 through 12, respectively, for performance analysis and results interpretation. The angular ranges for azimuth and elevation are $[-180^\circ, 180^\circ]$ and $[-45^\circ, 45^\circ]$, respectively. 

We follow the cross-validation setup provided with the dataset, that has 400 recordings for training, 100 for validation and 100 for testing. The incremental learning uses the 400 recordings in training by considering only the 8 target classes in Stage 0. The other 4 classes, while possibly present in the audio, are considered at Stage 0 as out-of-class interfering sounds, similar to the other interfering sounds in the dataset. During training, we use the Adam optimizer with a learning rate of $1e^{-3}$. The model is trained for 100 epochs with a batch size of 128. The validation set is employed for model selection, while the test set is used for performance evaluation. The model's performance is evaluated based on its predictions and their correspondence to the true DOA estimates. 

\subsection{Evaluation Metrics}
For the performance analysis of SELD, we use spatially-thresholded Error Rate ($ER \leq T^\circ$) and F1-score ($F \geq T^\circ$) to assess detection performance, penalizing correct detections that fall outside a specified angular distance threshold from the reference DOAs. Similar to baseline, we set this threshold at $T=20^\circ$ \cite{mesaros2019joint,politis2022starss22}. In addition to these location-dependent metrics, we compute two localization metrics: Class-dependent Localization Error (LE) and Localization Recall (LR), which are calculated individually for each class and then averaged. LE represents the mean angular error, determined by pairing predicted DOAs with their closest reference DOAs, while LR indicates the true positive rate of detected localization estimates out of the total instances for a given class. 
LE and LR offer complementary insights into localization accuracy beyond the thresholded performance of the F1-score. All metrics are computed using one-second non-overlapping frames. For a more detailed explanation of the joint SELD metrics, please refer to \cite{politis2020overview, mesaros2019joint}.

\subsection{Baseline methods}
We compare our proposed approach with the following methods that we consider baseline methods:
\begin{itemize}
    \item  We use the SELD baseline model from the DCASE 2022 Challenge configured for a single ACCDOA setting \cite{politis2022starss22}. This baseline model is trained on the complete dataset containing all 12 sound classes and evaluated on this same dataset.   The model minimizes the MSE loss between its predictions and the ground truth. This model is expected to show the best overall performance across all 12 classes since it learns all the classes at the same time.
    \item  Fine-Tuning (FT): In this method, the model from Stage 0 (trained on 8 classes) is fine-tuned to recognize all 12 classes. However, during fine-tuning, the model has access to training data consisting only of new classes, while still being evaluated on all 12 classes. The loss function minimizes the MSE between predictions and ground truth for all 12 classes. FT often leads to catastrophic forgetting, as the model tends to forget previously learned classes while learning the new classes because the data used for FT does not contain any examples of the previous classes.  
\end{itemize}
\section{Results and Discussions}
Table \ref{tab:my-table} compares overall SELD performance across all 12 classes for each method. For completeness, we include results for $\lambda = 0$, where the system is trained at the incremental stage using only $\mathcal{L}_{\text{MSE}}$. This setting results in an Independent Learning (IndL) approach, in which the model at Stage 1 is trained exclusively on the 4 new classes \cite{mulimani2023incremental}. Unlike FT, IndL minimizes the MSE loss only for the new classes without accounting for outputs related to the previously learned classes. The proposed CIL-MSE employs a balanced loss (with $\lambda = 0.5$) as defined in \ref{eq:1}. We also include results using Kullback-Leibler divergence instead of MSE for the distillation loss $\mathcal{L}_{\mathcal{OD}}$. 

The baseline SELD model, trained on the full dataset, achieves the best results with balanced localization and detection metrics. 
The proposed CIL-MSE and CIL-KLD methods ($\lambda = 0.5$), which combine new class learning and distillation-based retention, perform similar to the Baseline by striking a balance between old and new knowledge. CIL-MSE achieves better overall metrics than CIL-KLD, probably due to its MSE-based distillation loss which directly minimizes discrepancies in logits. In contrast, FT and IndL struggle with catastrophic forgetting, showing a steep drop in performance with high LE and low F1-scores, caused by the forgetting of previously learned classes. 
\begin{table}
\centering
\caption{Performance metrics computed for different methods averaged across all 12 classes. CIL-*:proposed class-incremental learning}
\label{tab:my-table}
{%
\begin{tabular}{|l|c|c|r|r|l|r|}
\hline\textbf{Model} & $\mathcal{L}_{\text{MSE}}$ & $\mathcal{L}_{\mathcal{OD}}$ & \textbf{LE} $\downarrow$   & \textbf{LR} $\uparrow$ & \textbf{ER} $\downarrow$ & \textbf{F1} $\uparrow$ \\ \hline

Baseline & - & - & 39.4 & 41.2 & 0.70 & 23.1 \\ \hline

\begin{tabular}[c]{@{}l@{}} FT \end{tabular}  &  \checkmark & - & 127.4 & 12.5  & 0.80 & 9.4 \\ 

\begin{tabular}[c]{@{}l@{}} IndL \end{tabular} & \begin{tabular}[c]{@{}l@{}}  \checkmark  \end{tabular} & - & 96.1 & 16.0 & 0.90 & 10.5 \\ \hline



\begin{tabular}[c]{@{}l@{}} CIL-MSE \end{tabular} & \begin{tabular}[c]{@{}l@{}} \checkmark \end{tabular} & \checkmark & 26.6 & 40.9 & 0.71 & 23.1 \\ 

\begin{tabular}[c]{@{}l@{}} CIL-KLD\end{tabular} & \begin{tabular}[c]{@{}l@{}} \checkmark \end{tabular} & \checkmark & 25.9  & 34.5  & 0.76 & 20.0   \\ \hline 
\end{tabular}%
} 
\end{table}
\begin{table}
\centering
\caption{F1-scores computed separately for old classes, new classes, and averaged score after the incremental phase.}
\label{tab:my-table1}
{%
\begin{tabular}{|l|c|c|c|}
\hline
Method & $C_{old}$ (1-8) & $C_{new}$ (9-12) & Overall (1-12) \\ \hline
FT & 0.0 & 28.2 & 9.4 \\ 
IndL & 0.3 & 31.0 & 10.5 \\ \hline
CIL-MSE & 19.7  & 29.8 & 23.1 \\ 
CIL-KLD &  21.2 & 17.7 &  20.0\\ \hline
\end{tabular}%
}
\end{table}

Table \ref{tab:my-table1} provides in detail the F1-scores averaged across old (Stage 0) and new (Stage 1) classes to highlight the performance differences between classes learned at each stage. FT and IndL favor learning the new classes but completely forget the old ones, leading to low overall F1 scores. The proposed CIL methods maintain good F1 scores across both old and new classes. The CIL-KLD method slightly outperforms CIL-MSE on old classes due to its KLD-based loss, which aligns probability distributions and seems to preserve prior knowledge more effectively. Notably, CIL-MSE excels with new classes, as its MSE-based loss penalizes large logit errors, supporting finer alignment with previous predictions. This approach effectively preserves key characteristics of old classes while allowing flexibility to learn the new ones.
 
 \subsection{Class-wise Analysis}
 
 \begin{figure*} [!t]
    \centering
    \begin{subfigure}{0.49\textwidth}
        \centering
        \includegraphics[width=\textwidth]{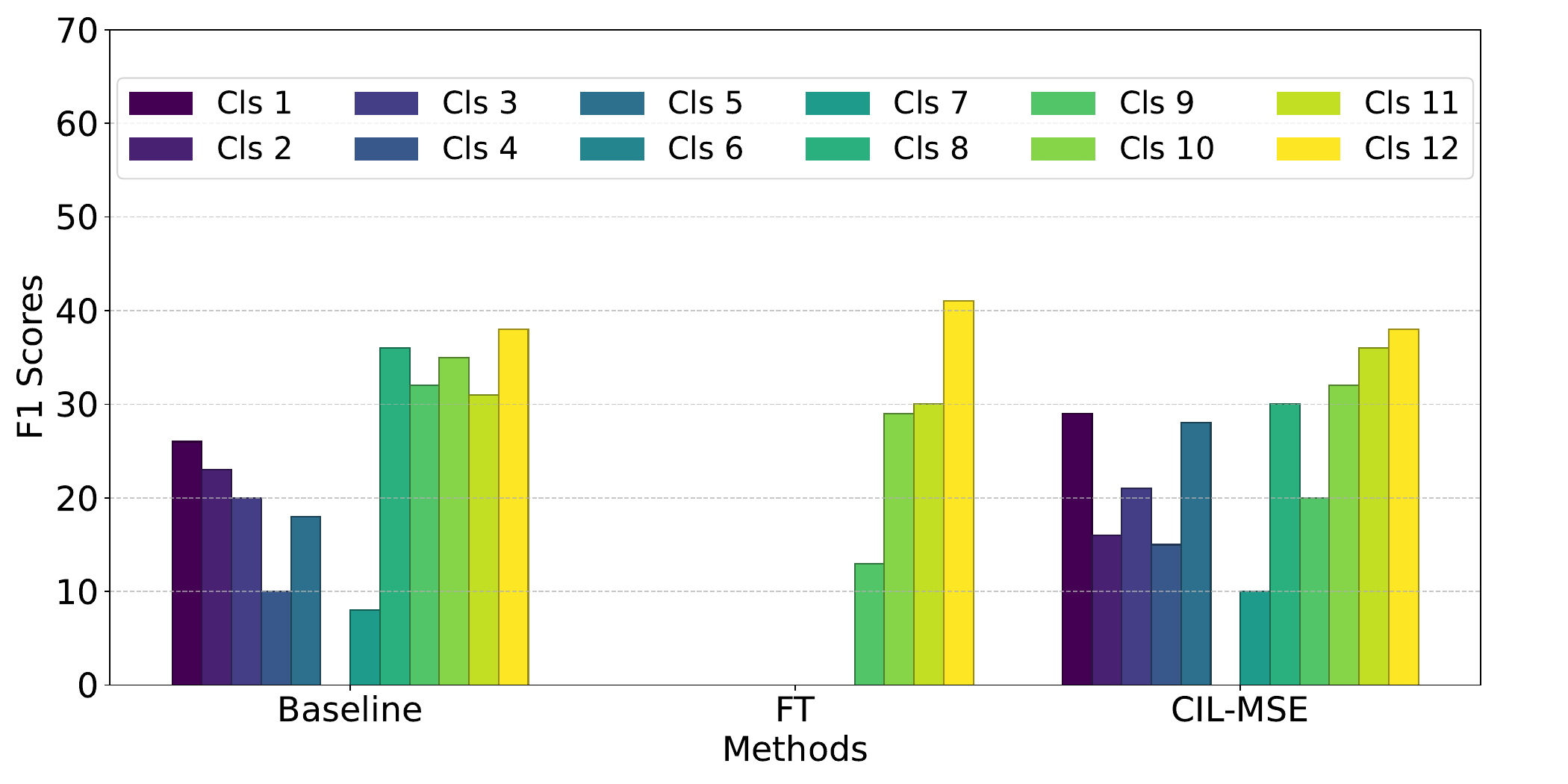}
        \caption{F1 scores for all 12 classes}
    \end{subfigure}
    \hfill
    \begin{subfigure}{0.49\textwidth}
        \centering
        \includegraphics[width=\textwidth]{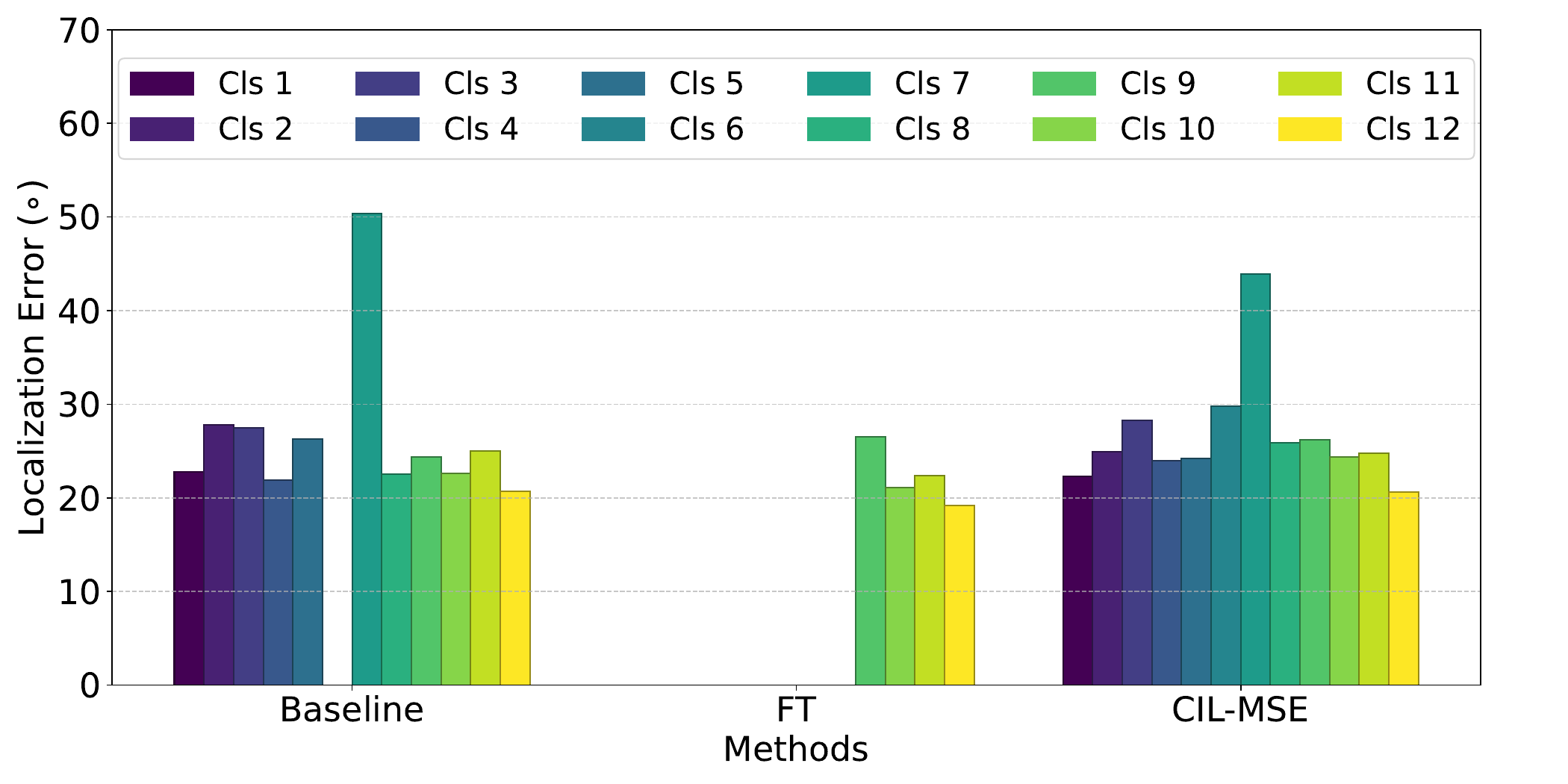}
        \caption{LE for all 12 classes}
    \end{subfigure}
    \caption{Class-wise F1-scores and localization error for different methods.}
    \label{fig:classwise_barplot}
    \vspace{-0.2cm}
\end{figure*}
Figure \ref{fig:classwise_barplot} presents the class-wise F1-scores and LE showing each method's efficacy in balancing performance across both old and new classes within the incremental learning setup. The Baseline model, trained on all 12 classes simultaneously, achieves generally strong performance across most classes but shows weaknesses in specific classes (e.g., 4 and 6), due to classes being difficult to recognize. FT exhibits catastrophic forgetting with zero F1 scores for old classes (1–8) but excels in new classes (9–12), particularly in classes 10, 11, and 12, as it focuses exclusively on learning new information. In the FT group we omitted the bars for LE corresponding to the classes which have zero F1-scores, as LE for these cases is $180^\circ$, showing that the system does not detect them at all. In contrast, CIL-MSE achieves a balance across both class groups showing higher adaptability in new classes and achieving strong F1 scores and low LE for both old and new categories.
\subsection{Effect of $\lambda$}
\begin{figure}
   \centering
	\includegraphics[width=0.49\textwidth]{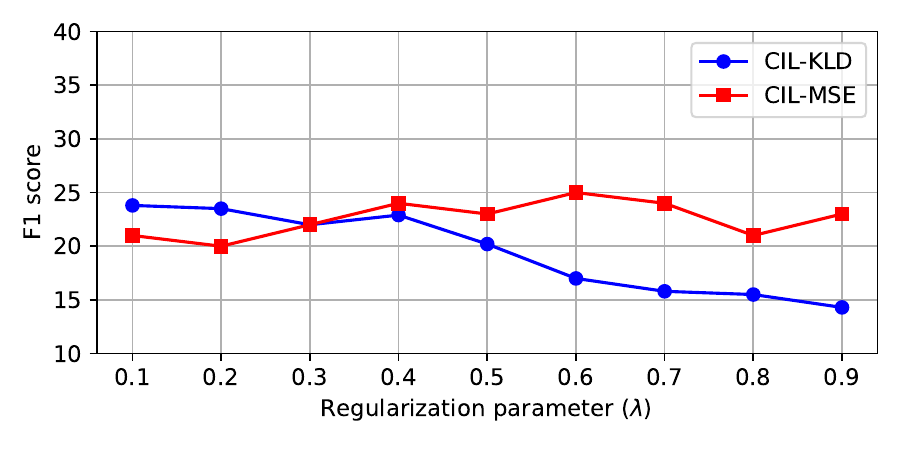}
	\caption{F1 scores of CIL-MSE and CIL-KLD for different $\lambda$}
	\label{fig:lambda_vs_f1score}
\end{figure}

Figure \ref{fig:lambda_vs_f1score} shows the average F1 scores across all 12 classes for different values of $\lambda$ in CIL-MSE and CIL-KLD. The plot shows distinct trends influenced by the balance between the new loss component $\mathcal{L}_{\text{MSE}}$ and the distillation loss $\mathcal{L}_{\mathcal{OD}}$ as controlled by varying $\lambda$. For CIL-KLD, the F1 scores decrease as $\lambda$ increases, indicating that a heavier emphasis on the distillation loss compromises the model's ability to adapt and learn new classes, resulting in lower overall performance. In contrast, CIL-MSE shows a more stable trend, peaking around $\lambda$ values of 0.5 to 0.6, indicating an effective balance between learning new classes and retaining prior knowledge. The smoother performance of CIL-MSE implies that its MSE-based distillation loss better manages class relationships during incremental learning, maintaining higher F1 scores with minimal degradation. Overall, CIL-KLD's declining trend shows the trade-off between retaining old knowledge and learning new information, while CIL-MSE shows greater adaptability and balance, highlighting the importance of loss function design in continual learning settings. 

\subsection{Analysis of regression weight vectors}
\begin{figure} [!h]
    \centering
    \begin{subfigure}{0.24\textwidth}
        \centering
        \includegraphics[width=\textwidth]{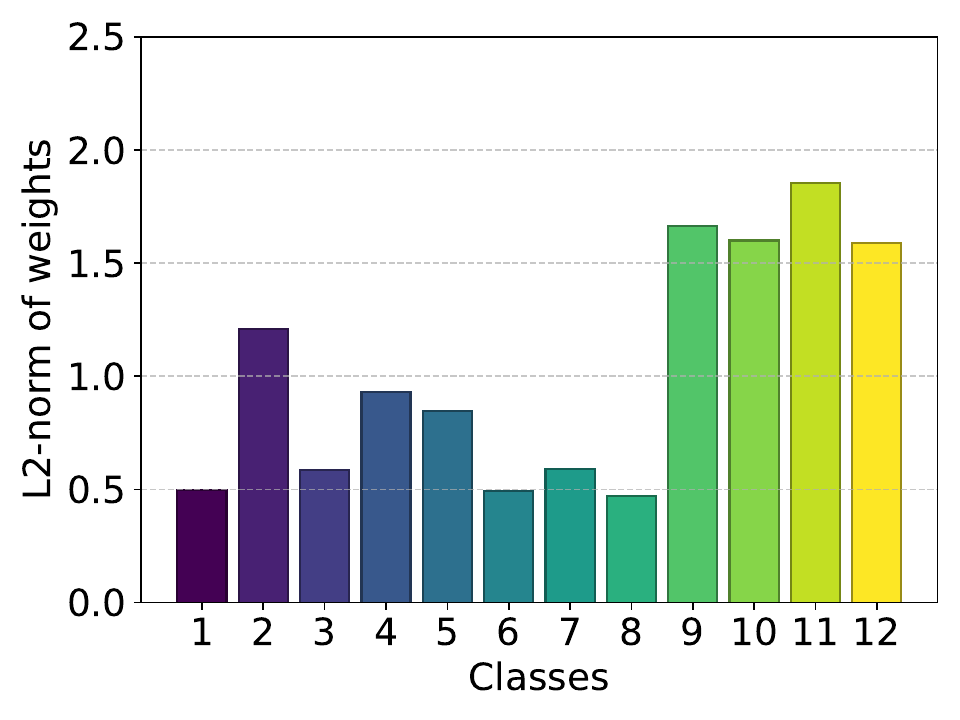}
        \caption{Fine Tuning}
    \end{subfigure}
    \hfill
    \begin{subfigure}{0.24\textwidth}
        \centering
        \includegraphics[width=\textwidth]{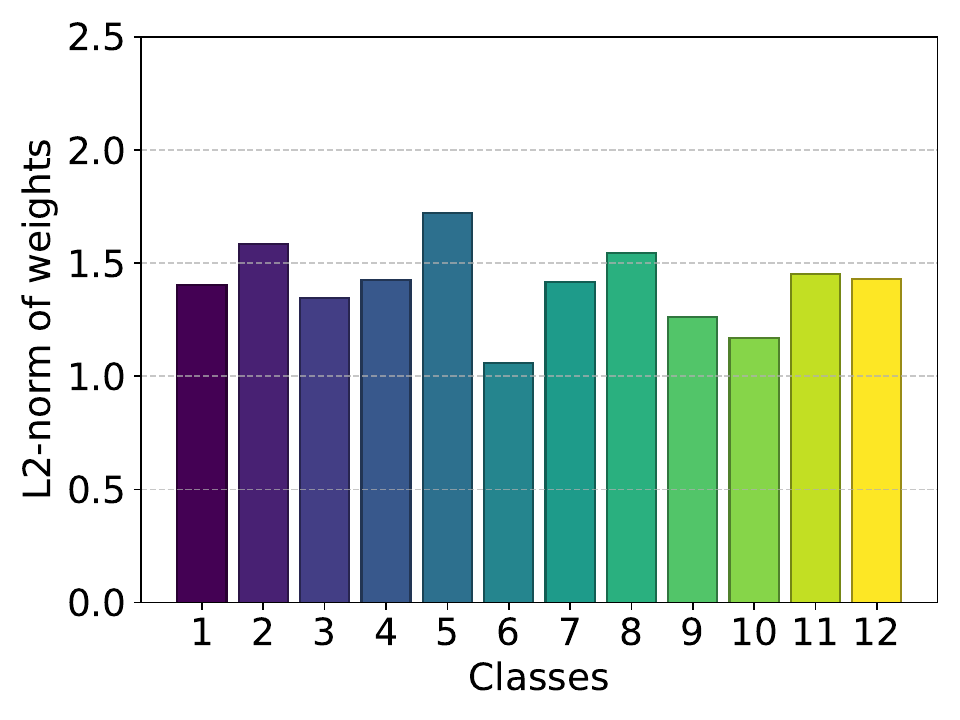}
        \caption{CIL-MSE}
    \end{subfigure}
    \caption{L2-norm of the regression weight vectors}
    \label{fig:l2_norm_weights}
\end{figure}
The final layer's class-specific weights provide insight into how the model handles the knowledge about the classes, serving as an indicator of balanced learning and stability in performance across all classes. Figure \ref{fig:l2_norm_weights} presents a comparison of the L2-norms of regression weight vectors for FT and CIL-MSE at Stage 1. In FT, the L2-norm is notably higher for the 4 new classes while substantially lower for the previous 8 classes, indicating the model’s tendency to focus on new classes and forget prior ones, even with weight initialization from Stage 0. In contrast, CIL-MSE maintains a balanced L2-norm distribution across all 12 classes, demonstrating its ability to integrate new class information without erasing prior knowledge. This balanced norm distribution underscores CIL-MSE’s capability in balancing distillation and new loss terms, which is essential for stable incremental learning in dynamic acoustic settings.
\section{Conclusions}
\label{sec:conclusions}
In this study, we proposed an incremental learning model for SELD systems designed to learn new classes independently while preserving knowledge of previously learned ones. We introduced an MSE-based output distillation loss to maintain information from earlier stages, and experimental results indicate that CIL-SELD performs similar to the baseline, showing improved localization for both existing and newly introduced classes without training on entire dataset. Although here we consider CIL-SELD in a two-stage learner setup, the model can be easily extended to multiple stages, depending on the total number of primary model classes and the number of classes introduced in each incremental step. Future work will explore multiple incremental steps and the most recently introduced distance estimation to further enhance localization capabilities, supporting the development of more adaptable and robust SELD systems for real-world applications.

\bibliographystyle{IEEEtran}
\bibliography{refs}


\end{document}